\documentclass[prl,aps,twocolumn,superscriptaddress,showpacs]{revtex4}

\usepackage{amsbsy}
\usepackage{amssymb}
\usepackage[dvips]{graphicx}
\usepackage{psfrag}

\newcommand {\V} {V} 
\newcommand {\Tc} {0.6514}

\begin{document}

\title{The mean field infinite range 
$p=3$ spin glass: equilibrium landscape and
correlation time scales}

\author{Alain Billoire}
\affiliation{
  Service de Physique Th\'eorique,
  CEA Saclay,
  91191 Gif-sur-Yvette, France.}

\author{Luca Giomi}
\affiliation{
  Dipartimento di Fisica, SMC and UdR1 of INFM and INFN, 
  Universit\`a di Roma {\em La Sapienza},
  P. A. Moro 2, 00185 Roma, Italy.}
\affiliation{
  Physics Department, Syracuse University, Syracuse NY 13244-1130, USA.}

\author{Enzo Marinari}
\affiliation{
  Dipartimento di Fisica, SMC and UdR1 of INFM and INFN, 
  Universit\`a di Roma {\em La Sapienza},
  P. A. Moro 2, 00185 Roma, Italy.}

\date{\today}                                                 

\begin{abstract}
We investigate numerically the dynamical behavior of the mean field
3-spin spin glass model: we study equilibrium dynamics, and compute
equilibrium time scales as a function of the system size $\V$.  We
find that for increasing volumes the time scales $\tau$ increase like
$\ln \tau \propto \V$. We also present an accurate study of the
equilibrium static properties of the system.
\end{abstract}


\maketitle


\section{Introduction}

The glassy state is a state of matter ubiquitous in nature. It enjoys
a number of intriguing properties: among those is the amazingly slow
dynamics, common to both structural glasses and spin glasses.  It is
surely not well understood, and during the last years it has been the
subject of numerous investigations, both theoretical and experimental.

One interesting approach, first used for studying spin glasses 
in \cite{MACYOU}, is based on (optimized) Monte Carlo numerical
simulations. One considers samples that can be brought to equilibrium
(in order to do that the lattice size cannot be too large) and studies 
numerically the equilibrium dynamics of the system (this is typically
not what happens with real disordered systems of this type, 
that are forever out of equilibrium).

The method was first used to investigate the dynamics of the mean
field Sherrington--Kirkpatrick (SK) model \cite{MACYOU,BIMA}: it turns out
\cite{BIMA} that all the relevant time scales $\tau_\alpha$ of the
model grow for diverging  lattice sizes $\V$ according to the
scaling law $\ln \left(\tau_\alpha\right) \propto \V^{1/3}$.  A single
time scale controls the critical behavior of the model: the time scale
governing the mode leading from $\cal A$, one of the many equilibrium
states of the mean field theory, to a different state $\cal B$ (not
related to $\cal A$ by a global spin flip) has the same scaling law
that the time scale controlling the mode related to a $Z_2$ global
inversion (leading from a state to its $Z_2$ symmetric).

We will deal here with the case of the mean field, fully connected
$p$-spin spin glass, with $p = 3$, and try to determine the relevant
time scales. Recently Ioffe, Lopatin and Sherrington have developed
\cite{IOIOSH} a theoretical analysis of the problem: their approach
suggests that for this class of models the relevant time scale behaves
as $\ln \left(\tau_\alpha\right) \propto \V$ (see also the work of
\cite{BIBRKO} for a computation of relaxation times in spin systems
with disordered quenched couplings).  Here we try to analyze the model
to uncover the true asymptotic behavior by using the numerical
approach of \cite{MACYOU,BIMA}.

The fully connected $p$-spin model is a generalization of the SK mean
field spin glass to a model where interaction is given by products of
$p$ spins (the $p=2$ model coincides with SK): it turns out to be a
promising candidate to understand the physics of structural glasses
\cite{MEPAVI,KI}.  Numerical simulations of this model are extremely
CPU time and memory consuming: one needs to store
order of $V^p$ quenched random couplings, and the time needed to run a
Metropolis sweep increases with volume like $\V^p$, as compared to
$\V$ for short range models. Furthermore, glassy slow dynamics
has to be studied for very large times. Here why study the $p=3$ fully
connected model: it is already quite expensive, but the cheapest of its
class (we do not loose in generality since all models with $p\ge 3$
are believed to have the same universal behavior).

In the next section we give details about our numerical
simulations. We use parallel tempering \cite{PARTEM}, an optimized
Monte Carlo method very effective on spin glass systems \footnote{The
$3$-spin spin glass is computationally demanding, and its phase
transition has some of the features of a first order phase
transition\cite{MEPAVI}: methods like parallel
tempering have accordingly to been used with some additional care. We find that on
the lattice volumes we study the method is indeed performing in a
satisfactory way.}. As an added
bonus we present in the following section very precise results about the
statics of the system: these are by far the most accurate results
obtained  for equilibrium expectations in this model (see
\cite{PIRISA} for the former state of the art). They allow to compare
the $N=\infty$ analytic results with finite $N$ accurate values, and
to get a feeling about what a ``large'' lattice really is. They also
help in making us confident that we are really reaching
thermodynamic equilibrium.  The evidence collected here is
interesting since, although this model is very important for reaching
an understanding of the glassy phase, it has never been studied numerically with high
accuracy, because of  the extreme difficulty of
the simulation.

In a fourth section, we present our main results, about time scales in
the model. We find that the Ioffe--Lopatin--Sherrington approach leads
indeed to the correct estimate, and that $\ln \left(\tau_\alpha\right)
\propto \V$. We end by drawing our conclusions.

\section{Numerical Simulations}

The infinite range, fully connected, $3$-spin spin glass is defined by the
Hamiltonian
\begin{equation}
{\cal H}
\equiv
-{\displaystyle \sum_{i<j<k}J_{i,j,k}\ \sigma_i\, \sigma_j\,
  \sigma_k}\ ,
\end{equation}
where the quenched random exchange coupling $J_{i,j,k}$ govern the
interaction among triples of spins and can take one of the two values
$\pm \sqrt{3}/\V$ with probability one half.  The choice we made of 
binary couplings allows far superior performances of the computer
code.

The model has a complex phase diagram\cite{GARDNER,CAPARA,MORITE}.  From a
thermodynamic point of view it has three phases.  At high $T$ it is
in a paramagnetic replica symmetric (RS) phase. At $T=T_c$ the system
enters a one-step replica symmetry breaking (1RSB) phase, while at a
lower temperature value $T_G<T<T_c$ it enters a third phase that will
not concern us here.

The order parameter is the usual overlap $q$.
Its probability distribution function can be written as
\begin{equation}
P(q) \equiv
\frac{1}{\V}\overline{\left\langle
\delta(q-\sum_i \sigma_i \tau_j)\right\rangle}\ ,
\end{equation}
where, as usual, the brackets denote a thermal average and the
over-line denotes an average over the quenched random couplings.  In
the RS phase, in the infinite volume limit, two different replica have
zero overlap.  In the 1RSB phase two different replica have overlap
$q_0=0$ with probability $m$, and overlap $q_1>0$ with probability
$1-m$ (both $q_1$ and $m$ depend on temperature, and should be written
$q_1(T)$ and $m(T)$). The values of $m$ and $q_1$ are solutions of a
set of two coupled integral equations.  The condition $m=1$ gives the
value of the critical temperature $T_c \approx \Tc$ (the expansion
around the large $p$ limit of \cite{GARDNER} gives $T_c\approx
0.6671$, and $T_G\ \approx 0.24$). It turns out that $\lim_{T\to
T_c^-}q_1(T)\neq 0$, namely the model has discontinuous 1RSB.

\begin{figure}
  \centering {\psfrag{x}[c][c][0.75]{$T$}
  \psfrag{y}[r][r][0.75][-90]{$E(T)$}
  \includegraphics[width=0.33\textwidth,angle=270]{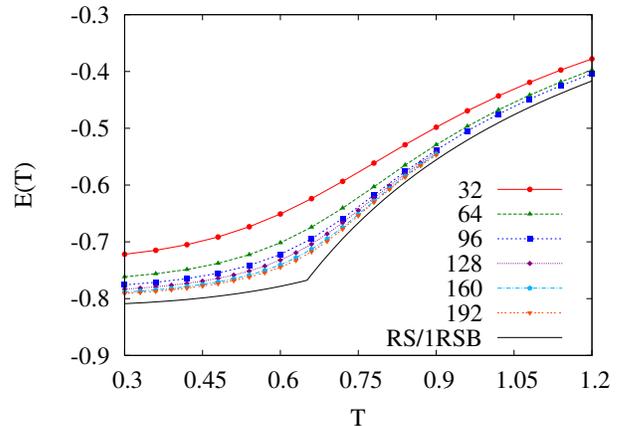}}
  \caption[a]{The internal energy as a function of the temperature. The
  continuous curve is from the analytical solution of the
  model. Numerical data are for system sizes from $32$ to $192$.}
\protect\label{en_color}
\end{figure}

\begin{figure}
  \centering {\psfrag{x}[c][c][0.75]{$T$}
  \psfrag{y}[r][r][0.75][-90]{$E(T)$}
  \includegraphics[width=0.33\textwidth,angle=270]{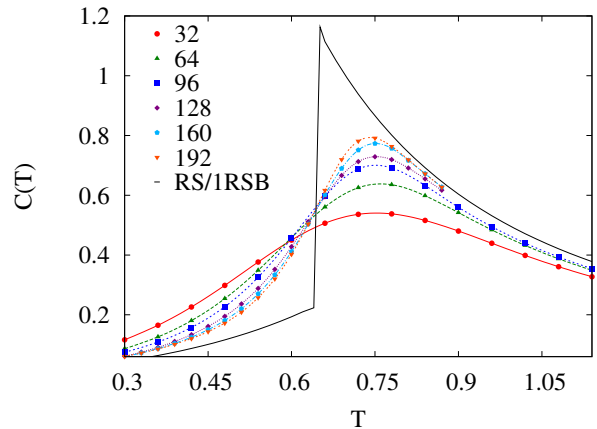}}
  \caption[a]{As in figure \ref{en_color}, but for the specific heat.}
\protect\label{cv_color}
\end{figure}

Let us start by giving some details about our simulation.  We study
systems with $\V=32$, $64$, $96$, $128$, $160$ and $192$ spins (the
needed CPU time increases with volume like $\V^3$).  We first
thermalize the system using the {\em parallel tempering} optimized
Monte Carlo procedure \cite{PARTEM}, with a set of $15$ $T$ values in
the range $0.3-1.2$ (i.e. $\Delta T = 0.06$) for the three smallest
volumes and with $\Delta T = 0.03$ in the temperature range  $0.3-0.9$
for the three largest volumes.  We take advantage of
the binary distribution of the couplings to use the multi-spin coding
technique\cite{MSC}, with a one order of magnitude gain in update
speed.  We perform $4\cdot 10^5$ iterations (one iteration consists of 
one Metropolis sweep of all spins plus one tempering update cycle), and
store the final well equilibrated configurations.

\begin{figure}
  \centering {\psfrag{x}[c][c][0.75]{$T$}
  \psfrag{y}[r][r][0.75][-90]{$\bar q(T)$}
  \includegraphics[width=0.33\textwidth,angle=270]{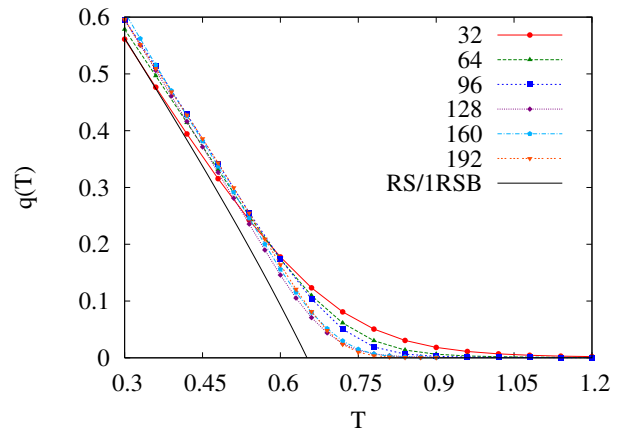}}
  \caption[a]{The order parameter $\bar{q}$ as a function of the
  temperature. The continuous curve is from the analytical solution of
  the model. Data are for system sizes from $32$ to $192$.} 
  \protect\label{q_color}
\end{figure}

We then, for studying the dynamical behavior of the system, start
updating these equilibrium configurations with a simple Metropolis
dynamics, and perform $2 \cdot 10^6$ Metropolis sweeps.  The number of
disorder samples is $1000$ for the $\V=32$ system, $500$ for $\V=64$
and $200$ for the other, larger, sizes.  As usual, the program
simulates the independent evolution of two replicas, in order to
compute the overlap $q$.  All statistical error estimates are done
using a jackknife analysis of the fluctuations among disorder samples,
with $20$ jackknife bins.

We have used the second half of the $4 \cdot 10^5$ thermal sweeps, i.e. the
last $2 \cdot 10^5$, to measure static quantities. This is on the one side an
interesting result on his own (since we can compare to the analytic,
$N=\infty$, result, and get hints about how finite $N$ corrections
work), and allows, on the other side, to check thermalization.  Notice
that our largest thermalized system is more than five times larger
than the largest analyzed before \cite{PIRISA}: we will present next
these results, before discussing the dynamical behavior of the system.

\section{Static Behavior}

\begin{figure}
  {\psfrag{x}[c][c][0.75]{$q$} \psfrag{y}[r][r][0.75][-90]{$P(q)$}
  \centering
  \includegraphics[width=0.33\textwidth,angle=270]{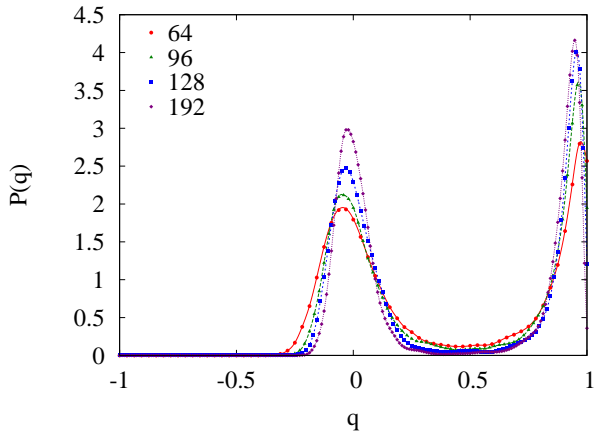}}
  \caption[a]{Probability distribution function of the overlap, for
  system sizes from $32$ to $192$. Here $T=0.42$.} 
\protect\label{pq_color}
\end{figure}

\begin{figure}
  \centering {\psfrag{x}[c][c][0.75]{$T$}
  \psfrag{y}[r][r][0.75][-90]{$B(T)$}
  \includegraphics[width=0.33\textwidth,angle=270]{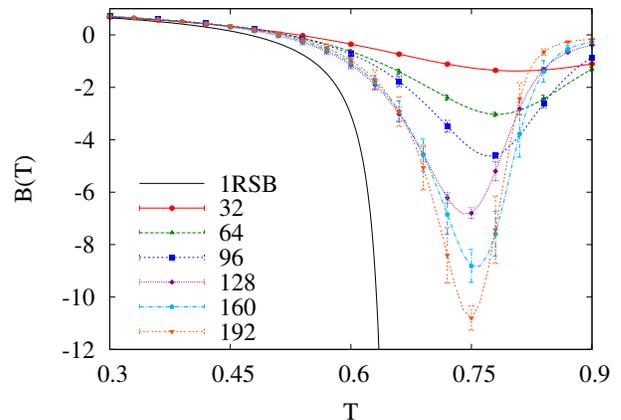}}
  \caption[a]{Binder parameter as a
  function of the temperature. The continuous curve is from the
  analytical solution of the model.}
  \protect\label{b_color}
\end{figure}

We show in figure \ref{en_color} our data for the internal energy 
$E ={\overline{\left\langle {\cal H}\right\rangle}}/{\V}$ 
as a
function of the temperature (we average this single replica quantity
over the two copies of the system that we follow in parallel),
together with the analytical result
\begin{equation}
 E = -\frac{1}{2T} \left(1-\left(1-m\right)q_1^3\right)\ .
\end{equation}
The finite volume numerical data converge nicely toward the
infinite volume limit analytical curve. This statement can be make
quantitative by fitting to some assumed analytical finite size
behavior.  Taking $T=0.3$ as an example, a good representation of the
data is obtained with the simple form
$E_{\V}=E_{\infty}+a V^{-1/\vartheta}$: by using for $E_{\infty}$ the
exact analytic value, one finds that a best fit gives 
$\vartheta=1.13\pm 0.01$.  
In figure \ref{cv_color} we show the specific heat: again, the
approach to the infinite volume result is very clear in the numerical
data. 

\begin{figure}
  \centering {\psfrag{x}[c][c][0.75]{$T$}
  \psfrag{y}[r][r][0.75][-90]{$A(T)$}
  \includegraphics[width=0.33\textwidth,angle=270]{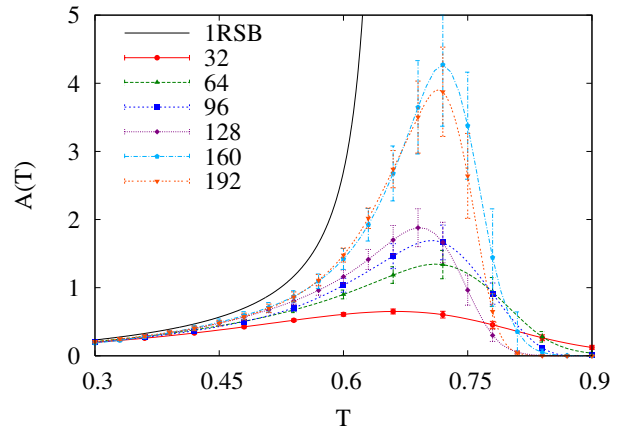}}
  \caption[a]{The $A$ parameter as a
  function of the temperature. The continuous curve is from the
  analytical solution of the model.  
} \protect\label{a_color}
\end{figure}

We show in figure \ref{q_color} the order parameter
$\overline{q}=\overline{\left\langle q\right\rangle}$, together with
the theoretical result $(1-m)q_1$, as a function of the temperature:
we find again an excellent convergence to the infinite volume limit.

We show in figure \ref{pq_color} the overlap probability distribution
(for $T=0.42$). Data are in agreement with the prediction of a bimodal
probability distribution, that becomes sharper and sharper as $\V$
grows: one peak is around zero and the other around some positive
value of $q$.  Between the two peaks the probability distribution
vanishes in the large volume limit, as predicted by the 1RSB picture
(and in contrast to $\infty$RSB).

Figures \ref{b_color} and \ref{a_color} are for the usual Binder
parameter
\begin{equation}
B\equiv
\frac{1}{2}\Bigl(3-\frac{\overline{\left\langle q^4\right\rangle}}
{\overline {\left\langle q^2\right\rangle}^2}\Bigr)\ ,
\end{equation}
and for the $A$ parameter that measures order-parameter fluctuations
\begin{equation}
A=\frac{\overline {\left\langle q^2\right\rangle^2}
-\overline{ \left\langle q^2\right\rangle}^2}
{\overline {\left\langle q^2\right\rangle}^2}\ .
\end{equation}
As shown in reference \cite{PIRISA}, $B$ and $A$ have very simple
expression
as a function of $m$, since for $q_0=0$ there is no
residual
dependence on $q_1$ and one gets that both $B$ and $A$ are equal to zero in the high $T$ phase, and that:
\begin{eqnarray}
B&=&\frac{2-3m}{2(1-m)} \qquad T_G<T<T_c\ ,\\
A&=&\frac{m}{3(1-m)} \qquad T_G<T<T_c\ .
\end{eqnarray}
Note that we are in a non-standard case where some dimensionless
quantities diverge at $T_c$: in this model both $B$ and $A$ diverge as
$T\to T_c^-$. This is in marked contrast with usual cases, where both
$B$ and $A$ have a finite limit for all temperatures as $\V\to\infty$.

Our numerical data are consistent with the predictions: namely they
show a zero limit in the RS phase, and a nonzero limiting curve, that
diverges as $T\to T_c$, in the 1RSB phase (data are consistent with
both the maximum of $A$ and the minimum of $B$ being proportional to
$\V$).  In other words, in this situation one finds, neither for $B$
nor for $A$, a fixed point where the curves for increasing values of
$\V$ cross.  Absence of crossing for the Binder parameter has been
also observed in \cite{HUKUKAWA} for the infinite range $3$ state
Potts spin glass model (a model with continuous 1RSB) and in
\cite{BIBRKO} for the $10$ state model (a model with discontinuous
1RSB). The fact that $A$ has a non-trivial limit as $\V$ grows shows
clearly that the model has non-zero order parameter fluctuations (OPF)
in the 1RSB phase, and the effectiveness of the parameter $A$ to
determine whether OPF holds or not.

\section{Dynamical Behavior}

\begin{figure}
  \centering
  {\psfrag{x}[c][c][0.75]{$t$}
  \psfrag{y}[r][r][0.75][-90]{$q^c(0,t)$}
  \includegraphics[width=0.33\textwidth,angle=270]{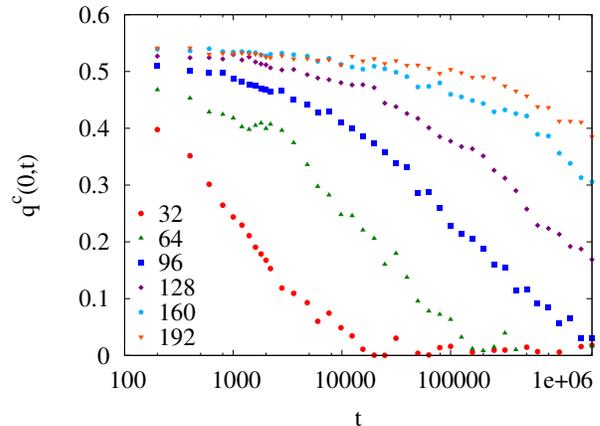}}
  \caption[a]{$q^c(0,t)$ as a function
  of $t$ (in log scale) in units of Metropolis sweeps. 
  Here $T=0.48$. }
\protect\label{q1c_0.48}
\end{figure}

The main result of this note is the precise quantitative measurement
of the typical time scales of the model and of their scaling
behavior. As we have already discussed we use the approach of 
\cite{MACYOU,BIMA}, considering the equilibrium dynamics of thermalized
configurations.

\begin{figure}
  \centering
  {
  \includegraphics[width=0.33\textwidth,angle=270]{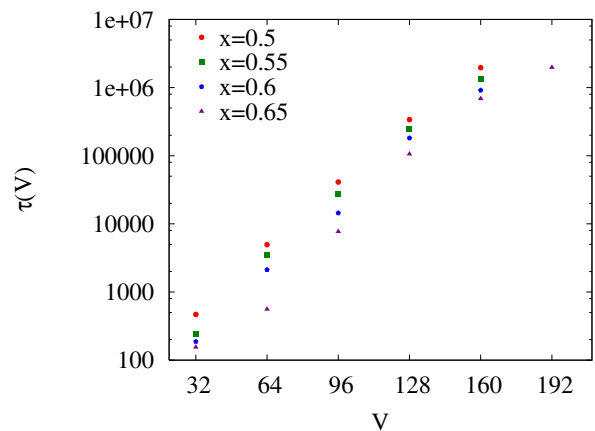}}
  \caption[a]{$\tau_X$ (in log scale) as a function of the number of
  spins $\V$, for $X=0.5$, $0.55$, $0.6$ and $0.65$. 
  $T=0.48$. } \protect\label{tau_x}
\end{figure}

We measure the time dependent overlap of two spin configurations:
\begin{equation}
q(0,t)\equiv
\left\langle \frac1N \sum_i \sigma_i(0) \sigma_i(t)
\right\rangle\ ,
\end{equation}
where we start, at time zero, from an equilibrium spin configuration
(see the former two sections), and the time $t$ is measured in units
of Metropolis sweeps (in this phase of the numerical simulation we are
using the simple Metropolis dynamics:  we use parallel
tempering only in the first phase of
thermalization and equilibrium analysis). The thermodynamic average $\left\langle
\dots\right\rangle$ is taken by averaging over several initial
configurations. As noticed in \cite{MACYOU}, the best and most
effective way to proceed is to use a new disorder sample for every simulation: we
do never repeat simulations with different starting points in a given
disorder sample, but always use new quenched random couplings when
starting a new numerical run.  This procedure correctly averages over
the thermal noise and over the random couplings, by minimizing the
statistical incertitude (mainly connected to the sample average): no
bias is introduced.  In what follows, the average $\left\langle
\dots\right\rangle$ is accordingly an average over two independent
replicas only.

As $t$ goes to infinity, $q(0,t)$ 
coincides with the static overlap, and it
is distributed according to the 
probability distribution $P(q)$. Accordingly (on a finite system)
\begin{equation}
q^c(0,t)\equiv
\overline{q(0,t)}-\bar{q} \to 0 \qquad as \ t \to \infty.
\end{equation}
Our results for $q^c(0,t)$ as function of the Metropolis time (note
the logarithmic scale) for $T=0.48$ can be found in figure
\ref{q1c_0.48} ($T_c\approx \Tc$, so we are at $T\sim\frac23 T_c$).
The behavior of $q^c(0,t)$ is very clear.  On the largest lattices we
are able to study, the time evolution has two successive regimes as $t$
grows: first $q^c(0,t)$ decays very slowly toward the value $q_1$
(the equivalent of $q_{EA}$ in spin glass models). It stays in this
first regime longer and longer as the system size increases (and would
stay there forever in the infinite system size limit). In a second
regime, $q^c(0,t)$ goes to zero, and the system equilibrates.

\begin{figure}
  \centering
  {\psfrag{x}[c][c][0.75]{$\V$}
  \psfrag{y}[r][r][0.75][-90]{$\tau_{1/2}$}
  \includegraphics[width=0.33\textwidth,angle=270]{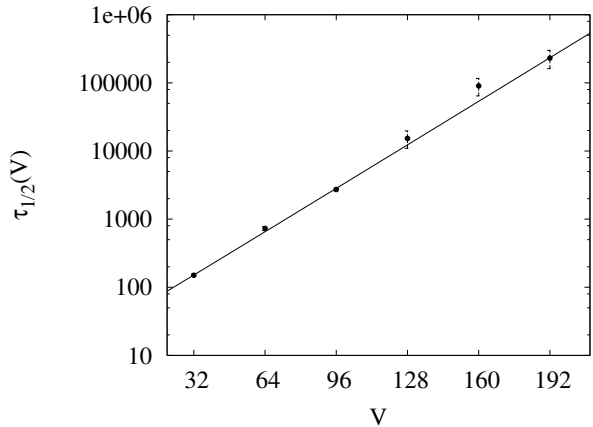}}
  \caption[a]{$\tau_{1/2}$ (in log scale) as
  a function of the number of spins $\V$. $T=0.54$. }
\protect\label{tau_0.54}
\end{figure}

We need now to define a typical time scale of the equilibrium
dynamics. We use the simple approach based on selecting a time such
that the (normalized) correlation decreases below a given threshold
$X$. We tried several values of this threshold to check the stability of
this definition. So our time scale $\tau_X$ is defined as the time
such that $q^c(0,t=\tau_X)=X\ q^c(0,0)$. Technically if the numerical
data cross the level $X$ more than once we average over all crossing
point (this is not a big issue, since we never find a real ambiguity,
but only sometimes a small wiggling, due to statistical fluctuations,
very local in time).  We now examine the values of $\ln \tau_X$ as a
function of $\V$.  If there is only one diverging time scale in the
problem, the behavior of $\tau_X$ as function of $\V$ should not
depend on $X$. We show the data obtained by this analysis in figure
\ref{tau_x} (for $T=0.48$): there is indeed a remarkable level of
universality, and the scaling law of the data is clearly  independent
of $X$, making us confident about the quality of our approach.
Notice that we are changing the threshold $X$ in a large range of
values, and the results are stable: this is a very good indication
toward the fact that we are really determining the physical time
scales.

In figure \ref{tau_0.54} we show our data for $\tau_{1/2}$ at
$T=0.54$, together with the best fit of the data to the expression
\begin{equation}
\tau\left(\V\right) = A \ e^{B \V^\epsilon}\ ,
\end{equation}
with the values $A=35 \pm 10$, $B=\left(4.5 \pm 2.0\right)\cdot
10^{-2}$ $\epsilon = 1.0 \pm 0.1$, with a value of chi-square
$\chi^2=1.8$ per degree of freedom.

These numerical results are in remarkable agreement with the
predictions of \cite{IOIOSH}: in the broken symmetry phase the typical
and relevant time scales grow here according to $\ln(\tau) \propto V$,
as opposed to the Sherrington--Kirkpatrick, $p=2$ model, where they
grow according to $\ln(\tau)\propto V^{\frac13}$ \cite{BIMA}.  Here
the absence of visible sub-leading corrections is remarkable, in
contrast again to the situation of the Sherrington--Kirkpatrick model
where the small exponent $1/3$ makes things difficult.

\section{Conclusions}

We have studied the $3$-spin infinite range spin glass model. We have
determined with good accuracy equilibrium properties on lattices of
reasonable size, gaining in this way an accurate control of finite
size effects.  The main point of this note has been to investigate the
equilibrium dynamics of the model, and to establish that the time
scales of the model grow with system size $\V$ according to the law
$\ln \tau_\alpha \propto \V$: this is in nice agreement with the
prediction of the theoretical approach by \cite{IOIOSH}.
Technically, from the computational point of view, these results are
worthy because simulating a $3$-spin infinite range model is a very
non-trivial task: here the computer time needed to perform a full
update of the lattice increases as $\V^3$ for large volumes.
We have succeeded on the one side to study the statics of the model on
lattice more than five times bigger than the largest ones used before,
and on the other side to determine with good accuracy severely
increasing time scales: we consider that as a worthy achievement.

\section{Acknowledgments}

We acknowledge enlightening discussions with David Sherrington in Les
Houches and on Montagne Sainte-Genevi\`eve.  We are especially
grateful to Giorgio Parisi for pointing out to us a series of
typographical misadventures that plagued a first version of the text.
The numerical simulations have been run on the PC cluster Gallega at
Saclay.

\end{document}